\documentclass[conference,10pt]{IEEEtran}
\usepackage{amsmath}
\usepackage{breakcites}
\usepackage{xspace}
\usepackage{graphicx}
\usepackage{diagbox}
\usepackage{url}
\usepackage{tabularx}
\usepackage{array}
\usepackage{booktabs}
\ifCLASSOPTIONcompsoc
  \usepackage[caption=false,font=normalsize,labelfont=sf,textfont=sf]{subfig}
\else
  \usepackage[caption=false,font=footnotesize]{subfig}
\fi
\usepackage{flushend}

\usepackage[hyperindex,breaklinks]{hyperref}
\usepackage{breakurl}

\newcommand{\Oracle}{\textsc{Astraea}\xspace}
\newcommand{\U}{\text{\O}\xspace}
\newcommand{\true}{\texttt{true}\xspace}
\newcommand{\false}{\texttt{false}\xspace}
\newcommand{\unknown}{\texttt{unknown}\xspace}

\newcommand{\Yspace}{\vrule height 3.2ex depth 2ex width 0pt} 
\newcolumntype{Y}{>{\Yspace\centering\arraybackslash}X}

\newcommand{\lowp}[1]{\ensuremath{< 1 \times 10^{#1}}}

\begin{document}
\title{\Oracle: A Decentralized Blockchain Oracle}



%
\author{
    \IEEEauthorblockN{John Adler\IEEEauthorrefmark{1},
    Ryan Berryhill\IEEEauthorrefmark{1},
    Andreas Veneris\IEEEauthorrefmark{1}\IEEEauthorrefmark{2},
    Zissis Poulos\IEEEauthorrefmark{1}, 
    Neil Veira\IEEEauthorrefmark{1}, and
    Anastasia Kastania\IEEEauthorrefmark{3}}
    \IEEEauthorblockA{\IEEEauthorrefmark{1}Department of Electrical and Computer Engineering,
    University of Toronto \\
    \texttt{\{adler,ryan,veneris,zpoulos,nveira\}@eecg.toronto.edu}}
    \IEEEauthorblockA{\IEEEauthorrefmark{2}Department of Computer Science,
    University of Toronto}
    \IEEEauthorblockA{\IEEEauthorrefmark{3}Department of Informatics, 
    Athens University of Economics and Business\\
    \texttt{ank@aueb.gr}}
}


\maketitle

\begin{abstract}
The public blockchain was originally conceived to process monetary
transactions in a peer-to-peer network while preventing double-spending.
It has since been extended to numerous other applications including 
execution of programs that exist on the blockchain called ``smart contracts.''
Smart contracts have a major limitation, namely they only operate
on data that is on the blockchain.
Trusted entities called oracles attest to external data in order to bring
it onto the blockchain but they do so without the robust security guarantees that blockchains 
generally provide. This has the potential to turn oracles into centralized points-of-failure.
To address this concern, this paper introduces \Oracle, a decentralized 
oracle based on a voting game that decides the truth or falsity of propositions.
Players fall into two roles: voters and certifiers. 
Voters play a low-risk/low-reward role that is resistant to adversarial manipulation
while certifiers play a high-risk/high-reward role so they are required to play with 
a high degree of accuracy. This paper also presents a formal analysis of the
parameters behind the system to measure the probability of
an adversary with bounded funds being able to successfully manipulate the oracle's decision,
that shows that the same parameters can be set to make manipulation arbitrarily difficult---a desirable feature for the system. 
Further, this analysis demonstrates that under those conditions a Nash equilibrium exists where
all rational players are forced to behave honestly.
\end{abstract}


%
\IEEEpeerreviewmaketitle


\section{Introduction}\label{section:intro}

Blockchain technology was originally developed to handle monetary
transactions in the form of a digital
currency using a peer-to-peer network while solving the double spending
problem~\cite{nakamoto2008}.  In its original application, it allowed a set of
anonymous and mutually-distrusting entities to reach consensus on the order of
a set of such transactions.  In recent years this concept has 
been extended to other domains in a way that 
enables such parties to reach consensus on items such as the order of private
monetary transactions~\cite{zksnarks,cryptonote}, the responsibility for file
storage in a peer-to-peer network~\cite{siacoin}, the sequence of executed 
states in generalized
computations~\cite{ethereum-yellow}, and more. 

A major limitation of blockchain technology is its inability to interact
with the ``outside world''~\cite{oracle-coindesk,oracle-ieeespectrum}. For 
example, in Ethereum~\cite{ethereum-yellow} the network can
reach consensus on the outcome of a sequence of computations.  However,
these computations can only operate on data that is {\em on} the blockchain,
that is, a prior computation initiated by a user of Ethereum wrote the relevant
data into the virtual machine's memory.
The network does not natively support reaching consensus on the validity
of such data, only the fact that it exists on the blockchain can be agreed upon.
As such, trusted entities called
\textit{oracles} are needed to attest to facts in an effort to
bring external data into the
consensus mechanism of a
blockchain.  Generally speaking, such oracles do not provide the robust security
properties of native blockchain protocols.  This may not be a substantial
limitation when the external data can be proven correct either
cryptographically (such as data that comes from another
blockchain~\cite{kiayias2017,herlihy2018}), or computationally (such as the
outcomes of complex computations that are infeasible to perform on platforms such
as 
Ethereum~\cite{truebit}). 
However, facts that cannot be provably verified like the ones above
present unique challenges for existing oracles.

TLSnotary~\cite{tlsnotary} and TownCrier~\cite{town_crier} are 
oracles that provide cryptographically-checkable information as they 
 attest to the content of websites accessed using the
Hypertext Transfer Protocol Secure (HTTPS) protocol. The underlying idea behind
these oracles is that attestations are checkable due to the use of transport layer security (TLS).
However, TLS does not guarantee that every visitor to the site sees the 
same information.
Additionally, a website could maliciously alter its output to influence on-chain
attestations, creating a centralized point-of-failure in the system. 
TownCrier uses Intel Software Guard Extensions in order
to protect the attestations against a malicious operating system.
In the case of Augur~\cite{augur}, it uses an oracle
to determine the outcome of events so as to pay out
participants in a prediction market.
The oracle may require specific users to report outcomes at specific times
or face a monetary penalty. As such, users
who maliciously report incorrect results may be subject to 
a dispute process.
While offering greater decentralization, Augur's oracle does not
offer users the opportunity to drop in and out of the system 
at will which is somewhat harmful to the usability of the ecosystem.

In this work, we propose \Oracle---a general-purpose decentralized oracle 
that runs on a public ledger and leverages human 
intelligence through a voting-based game. Without loss of generality, we
describe our contributions using the Ethereum blockchain but other 
platforms with similar smart contract characteristics can be utilized
as the underlying decentralized network. 
Entities in \Oracle can fall into one or more of three roles:
submitters, voters, and certifiers.
Submitters submit Boolean propositions to the system and pay
fees for doing so.
Voters play a low-risk/low-reward game where they are given the opportunity to
vote on the truth of a random proposition by placing a ``small''
stake in their confidence of their vote being correct.
Certifiers play a high-risk/high-reward game where they place a ``large'' 
stake on the outcome of the voting and certification process.
For usability, voters and certifiers are not required to be online
during particular time periods;
all roles may enter or exit the system at any time.
Due to its requirement for monetary staking and fees, 
the system is inherently not vulnerable 
to Sybil attacks~\cite{bentov2016}.

In more detail, in \Oracle
voters place a small amount of stake and are given the opportunity to
vote on a proposition selected randomly from the system.
This stake is deposited before seeing the proposition.
The outcome of voting is the stake-weighted sum of votes.
Due to the randomness involved when selecting a proposition, 
voting is resistant to manipulation by actors seeking to 
force an incorrect result for a particular proposition. On the other hand, 
certifiers place large stakes on the truth or falsity of propositions
of their choosing. 
Similar to voting, the outcome of certification is the
stake-weighted sum of certifications.
However, not all propositions necessarily carry certifications. Nevertheless, when
one exists and the voter outcome matches that of the certifier, 
players who took the corresponding position are
rewarded and players who took the opposing position are penalized.
When the voter and certifier outcomes disagree, 
all of the certifiers who took a position are penalized.
It is important to note that votes and certifications are sealed and revealed only
at the end of the aforementioned process.
Intuitively, the proposed system encourages certifiers to place bets on 
propositions for which there is a high degree of confidence that
they are true or false, and it also encourages voters
to vote accurately on the propositions they are randomly given.

Analysis reveals that the proposed system has desirable properties.
In this paper, we determine the probability of an adversary manipulating the voting 
outcome and we demonstrate that system parameters can be chosen in a way so that an 
adversary with a bounded amount of funds also 
has an arbitrarily small probability of successfully manipulating it.
Further, we demonstrate that under the same conditions that prevent an adversary
from manipulating voting, the system has a Nash equilibrium in which all players
play honestly.
Finally, we argue that the certification process avoids degenerate 
coordination strategies
in which users always vote with a constant ``true'' or ``false'' without
regard to the validity of the actual proposition so as to earn a profit.

The remainder of this paper is outlined as follows.
Section~\ref{section:preliminaries} gives a brief overview of blockchains, oracles, and key assumptions.
Section~\ref{section:related} follows with a summary of current blockchain oracle projects.
Sections~\ref{section:oracle} and~\ref{section:analysis} introduce \Oracle and analyze it from a game-theoretical direction.
An extension to \Oracle is presented in Section~\ref{section:extension}.
Finally, applications of a decentralized oracle are shown in Section~\ref{section:applications}.

\section{Preliminaries and System Assumptions}\label{section:preliminaries}

%
%
%
\subsection{The Distributed Public Ledger}

Bitcoin~\cite{nakamoto2008} is a blockchain-based distributed ledger
of monetary \textit{transactions} which transfers coins between users 
identified by their public keys and certified by digital signatures. 
Transactions are grouped into \textit{blocks} which are
ordered lists of transactions.
Each block contains a hash that commits to the previous block, 
forming a hash-linked chain (\textit{i.e.}, a blockchain).
An unambiguous head of the chain determines 
the precise order in which transactions occurred,
preventing \textit{double-spending}.
Bitcoin's~\cite{nakamoto2008} primary innovation was the use
of Proof-of-Work (PoW) to reach 
consensus on the head of the chain.
\textit{Miners} combine transactions into blocks.
Each block has a header consisting of a hash of its own transactions,
a hash of the previous block header, 
and a \textit{nonce} used for PoW.
Users select as the head the block with the most cumulative work in its history.
Rewriting the chain to undo a transaction requires repeating PoW,
and the further back in history the transaction is, 
the more work is required.

Ethereum~\cite{ethereum-yellow} uses similar ideas to generalize consensus
on the state of a virtual Turing-complete computer.
Ethereum transactions may additionally insert a 
\textit{smart contract}~\cite{szabo1997} into 
the virtual machine or call a function of an existing contract thus
extending the applicability and usability of a blockchain.

\subsection{Oracles}

This work focuses on a particular issue facing smart contracts: 
their inability to act on data that exists outside the blockchain.
Smart contract execution must be deterministic in order to be
publicly verifiable in perpetuity.
As such, they cannot directly pull data from \textit{e.g.}, the Internet.
For example, consider a contract to allow wagers on 
the outcome of the Mayweather vs McGregor boxing match~\cite{pmu_pmu_pmu}.
It would be straightforward to devise a smart contract that pays 
out bettors based on a flag representing the outcome of the match,
but trustlessly determining the value of that flag is a major challenge. 
In that sense, 
\textit{oracles} allow a blockchain to interact with the real,
outside world.

\subsection{System Assumptions}\label{section:preliminaries:assumptions}

This paper presents \Oracle, an oracle that decides the 
truth value of Boolean propositions.
We assume each proposition $p$ has a truth value $t$ that is 
either \true ($T$) or \false ($F$).
There are $N$ \textit{players} (voters and certifiers) numbered $1$ 
through $N$.
For each proposition $p$ and each $i \in [1,N]$, 
let $\beta_i(p) \in \{T,F\}$ denote the \textit{belief} of player $i$ 
regarding the truth value of proposition $p$.
Each player $i$ has an \textit{accuracy} $q_i \in [0,1]$ that is, informally, 
the probability that player $i$ is correct about a given proposition.
Formally:

\begin{equation}
 	\beta_i(p) = 
	\begin{cases}
		t      & \mbox{with probability } q_i       \\
		\neg t & \mbox{with probability } (1 - q_i) \\
	\end{cases}
\end{equation}

The value of $\beta_i(p)$ is independent of $\beta_j(p)$ for all $j \neq i$. 
That is, each player's beliefs are independent of all other players' beliefs.
Entities that have non-independent beliefs (\textit{e.g.}, users that pool their 
voting efforts in the style of mining pools~\cite{lewenberg2015}) 
are treated as a single player.
Further $\beta_i(p)$ is independent of $\beta_i(p')$ for all $p' \neq p$.
That is, a player's belief in a proposition is independent of her
beliefs in all other propositions.
An \textit{honest player} always votes or certifies that proposition
$p$ has truth value $\beta_i(p)$.

\section{Prior Art}\label{section:related}

The challenge of smart contracts interacting with the outside world has spawned numerous oracle proposals with varying degrees of centralization, performance, and scope tradeoffs.
A brief description of previous work is found here. 

\subsection{Existing Blockchain Oracle Proposals}

Hivemind (previously Truthcoin)~\cite{truthcoin} is a prediction marketplace built upon Bitcoin.
It allows users to report on the predicted outcome of a future event by staking \textit{reputation} currency, while traders trade on the event's market using base currency (\textit{i.e.}, Bitcoin).
However, it requires a complex scheme to balance
incentives involving two currencies and it has a permissioned and centralized
implementation due to token distribution limitations.
Augur~\cite{augur} is a prediction market platform built upon Ethereum, heavily inspired by Truthcoin.

Gnosis~\cite{gnosis} and its fork Delphi~\cite{delphi} 
are also prediction markets built upon Ethereum.
Both suffer from a gameable dispute resolution design and centralized token distribution.
ChainLink~\cite{chainlink} aims to provide a cross-chain portal to internet-available information \textit{i.e.}, data available on websites, through their centralized system.
Oraclize.it~\cite{oraclizeit} is a centralized platform that creates trusted blockchain transactions for use in smart contracts~\cite{pmu_pmu_pmu}.

Town Crier~\cite{town_crier} uses a stack of trusted hardware and software, and publishes proofs that this computational stack has not been tampered with.
This system is highly effective, so long as users trust that the
underlying hardware does not contain backdoors or exploits~\cite{meltdown,intel_amt}. 
In effect, the validity and accuracy of the protocol depends on a central authority.

All in all, the need of oracles to assist smart contracts was identified shortly after blockchain technology 
came into existence but no truly decentralized, trustless, and permissionless solution has been presented until this work.

\section{\Oracle}\label{section:oracle}


This section presents the decentralized voting-based oracle \Oracle.
It begins with a high-level overview of the user roles
and the operation of the voting game and concludes with a detailed
description of the game.
The next section analyzes the coordination game, illustrates its 
properties, and demonstrates
a desirable Nash equilibrium.

\subsection{Overview}

Users of \Oracle participate in one (or more) of the 
following three roles:
\textit{submitters}, \textit{voters}, and \textit{certifiers}, that
behave as follows.

\begin{figure}[tb]
\centering
\includegraphics{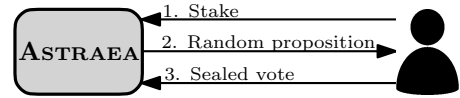}
\vspace*{-6pt}
\caption{Player voting in \Oracle}
\vspace*{-6pt}
\label{fig:voting}
\end{figure}

\begin{figure}[tb]
\centering
\includegraphics{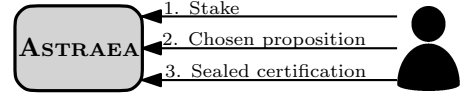}
\vspace*{-6pt}
\caption{Player certifying in \Oracle}
\vspace*{-12pt}
\label{fig:certifying}
\end{figure}

\begin{itemize}

\item \textbf{Submitters} 
choose which propositions enter into the system by
allocating money to fund (in part) the subsequent effort to validate the
Boolean propositions.

\item \textbf{Voters} play a low-risk/low-reward role.
Once such a player submits a deposit (\textit{stake}), she is given the chance 
to vote on a proposition chosen uniformly at random from the ones available.
The stake is placed before the voter is notified of which proposition she 
will be voting on.
The steps of this process are depicted in Figure~\ref{fig:voting}. 
The outcome of the voting process is a function of the sum of the votes 
weighted by the deposits.
The maximum voting deposit is a parameter of the system. 

\item \textbf{Certifiers} play a high-risk/high-reward role.
They choose an available proposition and place a large deposit in order to 
certify it as either true or false.
The certification outcome is a function of the sum of certifications weighted 
by the deposits.
This process is depicted in Figure~\ref{fig:certifying}.
Since certifiers choose which propositions to certify, not all
propositions are guaranteed to have certification, and indeed
the system does not mandate them to do so.
The minimum certification deposit size is a system parameter and should be 
large enough that certifying incurs a substantial risk.

\end{itemize}

Setting the parameters appropriately is important for the functionality of the
system.
The maximum voting deposit size should be small 
relative to the total voting stake on each proposition.
If a single vote can account for 100\% of a proposition's 
total voting stake, an adversary can have total control 
over the outcome of a randomly drawn proposition.
Conversely, if it is 1\%, the adversary would somehow need to draw
the same proposition repeatedly to control its validity outcome. 
On the other hand, the minimum certification deposit should be large enough that
certifiers incur sufficient risk.
Otherwise, they may be tempted to abuse their certifications for
griefing purposes as a malicious certification can impact reward payouts
for other players. 
At first sight, it seems like individual certifiers have enormous influence on
the process for individual propositions, and indeed this can be the case.
However, as described later in the paper, the certifiers alone cannot force
the oracle to produce an incorrect value and they are 
encouraged to behave honestly by the incentive structure; otherwise they
face large penalties.

\subsection{The Proposition List}

A key aspect of the system is the set of propositions over which
voters and certifiers play the game.
We assume that there exists a set of propositions constructed outside 
of \Oracle called the \textit{proposition list},
which is denoted by $P$ and is of fixed size $|P|$.
Each proposition $p_i \in P$ has a hidden truth value $t_i$ and is associated 
with a bounty amount $B_i$, the source of which is described later. 
The voting game described later is played over all propositions in $P$ simultaneously.
In addition, we assume there are two \textit{certifier reward pools}
containing $R_T$ and $R_F$ monetary units, intended to 
reward certifiers for \true
and \false outcomes, respectively.
The use of two separate reward pools is intended to avoid a degenerate 
coordination
strategy in which users always vote and certify with a constant
\true or \false so as to maximize their profit without expending any effort
and its rationale is justified later in this paper.

The proposition list can be constructed in a variety of ways and it is
outside the scope of this paper.
For instance, an auction could be conducted for the $|P|$ spaces,
with the auction prices becoming the bounty for each proposition.

After reaching a termination condition, propositions are
\textit{decided}, at which point the game outcome is computed, 
the proposition is removed from the list,
and the corresponding rewards and penalties are administered.
A proposition is decided after it has received a certain amount of voting 
stake denoted $D_v$, which is the same for all propositions.
The proposition can then be replaced with a new one by, for instance,
conducting another auction.

\subsection{System Description}

For each of the $N$ players, $s_{i,j,b}$ denotes the amount that player $i$ has 
staked on voting that proposition $p_j$ is $b$ where $b \in \{T,F\}$.
If player $i$ has not voted on proposition $j$, then 
$s_{i,j,T}$ and $s_{i,j,F}$ are both equal to $0$.
Otherwise, one or both (in the case where player $i$ has voted more than once 
in differing directions) are nonzero and equal to the stake with which
they voted.
Additionally, let $\sigma_{i,j,b}$ denote the amount of stake that
player $i$ used to certify that proposition $p_j$ is $b \in \{T,F\}$.
Let $s_{max}$ and $\sigma_{min}$ denote the maximum voting stake
and minimum certifying stake parameters, respectively.

\begin{figure}[tb]
\centering
\includegraphics[width=0.45\textwidth]{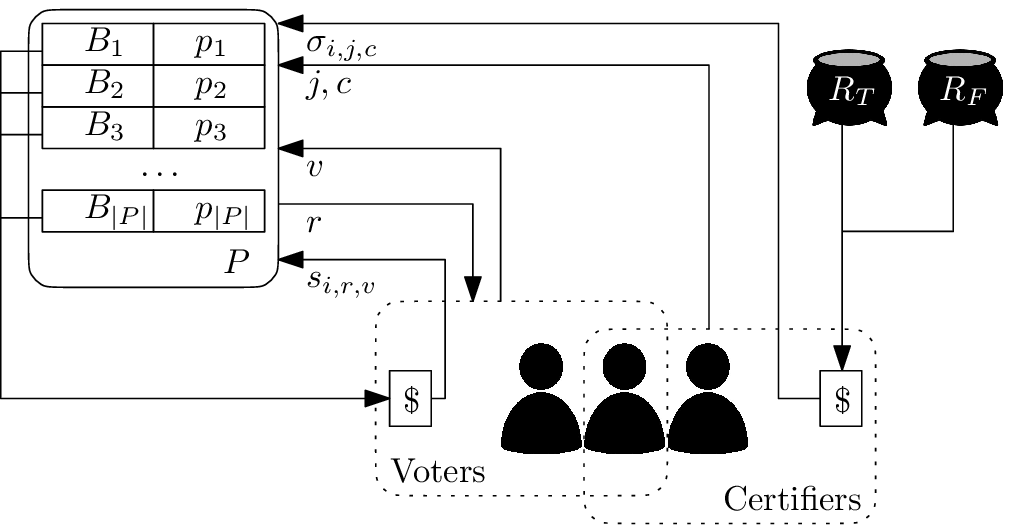}
\vspace*{-6pt}
\caption{Overview of \Oracle}
\vspace*{-12pt}
\label{fig:oracle}
\end{figure}

An overview of the game architecture is shown in Figure~\ref{fig:oracle}.
It depicts a set of players who are either voters, certifiers, or
both. In that figure, the differing mechanisms of interaction with the 
propositions can also be seen. 
Voters must engage in a multi-step process in order to vote on random 
propositions, while certifiers simply submit a certification to a 
proposition of their choosing in one step.

\subsubsection{Voting}
Players may place a deposit in order to vote on a randomly-chosen
proposition.
First, the voter indicates a desire to stake an amount of money
$s_{i,r,v} \le s_{max}$ on a vote for proposition $p_r$,
where $r$ is not yet known to the voter.
Subsequently, the system chooses the value for $r$ and sends it to the voter.
Note that $r$ is chosen uniformly at random from $[1,|P|]$, so
a voter may be given the opportunity to vote on the same proposition
more than once.
Generating random numbers in smart contracts is a topic of active research
and several techniques exist to do so securely~\cite{randao,micali1999},
including ones that use an oracle~\cite{oraclizeit}.
Finally, the voter submits a sealed vote $v \in \{T, F\}$. 
This can be accomplished using a commit-reveal scheme in which the voter
commits to a hash of their vote concatenated with a nonce and later reveals
the vote and nonce.
We define an \textit{honest voter} as one who always votes according to their belief $\beta_i(p_r)$.

\subsubsection{Certifying}
Players may place a large deposit in order to certify propositions
of their choosing.
The certifier simply submits a monetary stake $\sigma_{i,j,c} \ge \sigma_{max}$ 
and a sealed certification $c \in \{T, F\}$ for proposition $p_j$.
An \textit{honest certifier} is one who always certifies according
to their belief $\beta_i(p_j)$.

\subsubsection{Termination and Decision}

Once a proposition $p_j$ has accumulated sufficient voting stake, 
it is decided.
The amount of voting stake before decision is a parameter of the system
 denoted by $D_v$. 
Four values are computed:
$s_{TOT,j,T}$, $s_{TOT,j,F}$, $\sigma_{TOT,j,T}$, and $\sigma_{TOT,j,F}$.
These values represent, respectively, the total voting stake for \true,
total voting stake for \false, total certifying stake for \true,
and total certifying stake for \false.
For each $b \in \{T, F\}$, they are computed as follows:
\begin{align*}
    s_{TOT,j,b} = \sum_{i=1}^N s_{i,j,b} \qquad
    \sigma_{TOT,j,b} = \sum_{i=1}^N \sigma_{i,j,b}
\end{align*}

\begin{table}
\caption{Voting outcomes}
\vspace*{-4pt}
\label{tbl:vote_outcomes}
    \centering
    \begin{tabular}{|c|c|}
\hline \textbf{Outcome}  & \textbf{Condition} \\
\hline  $\mathrm{T}$     & $\mathrm{s_{TOT,j,T} > s_{TOT,j,F}}$ \\
\hline  $\mathrm{F}$     & $\mathrm{s_{TOT,j,F} > s_{TOT,j,T}}$ \\
\hline  \U               & $\mathrm{s_{TOT,j,T} = s_{TOT,j,F}}$ \\
\hline
    \end{tabular}
\vspace*{-8pt}
\end{table}

\begin{table}
\caption{Certifying outcomes}
\vspace*{-4pt}
\label{tbl:cert_outcomes}
    \centering
    \begin{tabular}{|c|c|}
\hline \textbf{Outcome}  & \textbf{Condition} \\
\hline  $\mathrm{T}$     & $\mathrm{\sigma_{TOT,j,T} > \sigma_{TOT,j,F}}$ \\
\hline  $\mathrm{F}$     & $\mathrm{\sigma_{TOT,j,F} > \sigma_{TOT,j,T}}$ \\
\hline  \U               & $\mathrm{\sigma_{TOT,j,T} = \sigma_{TOT,j,F}}$ \\
\hline
    \end{tabular}
\vspace*{-8pt}
\end{table}

Voting and certifying outcomes are computed as shown in 
Tables~\ref{tbl:vote_outcomes} and~\ref{tbl:cert_outcomes}.
It can be seen that a simple majority determines the outcome.
In case of a tie, the voting outcome is \unknown (denoted by \U).
The same formula applies for certification outcomes.
Alternatively, the system can require a super-majority instead of a simple
majority for a definite $T$ and $F$ outcome and produce a \U outcome
otherwise.
Ultimately this does not affect the design or the analysis of the game
and it is excluded from the description for the sake of simplicity.

The game and oracle outcomes that correspond to each of the nine 
possible combinations of
certification and voting results are illustrated in Table~\ref{tbl:outcomes}.
The headings in the top row correspond to certification, while
the labels in the first column correspond to voting.

\begin{table}
\caption{Outcomes for (a) the game and (b) the oracle}
\label{tbl:outcomes}
\parbox{0.45\linewidth} {

    \centering
    \begin{tabular}{|c|c|c|c|}
\hline      (a)             & \multicolumn{3}{|c|}{\textsc{Game}} \\      
\hline \backslashbox{V}{C}  & $\mathrm{T}$  & $\mathrm{F}$  & \U          \\
\hline  $\mathrm{T}$        & $\mathrm{T}$  & \U            & \U          \\
\hline  $\mathrm{F}$        & \U            & $\mathrm{F}$  & \U          \\
\hline  \U                  & \U            & \U            & \U          \\
\hline
    \end{tabular}
}
\hfill
\parbox{0.45\linewidth} {

    \centering
    \begin{tabular}{|c|c|c|c|}
\hline       (b)            & \multicolumn{3}{|c|}{\textsc{Oracle}} \\      
\hline \backslashbox{V}{C}  & $\mathrm{T}$  & $\mathrm{F}$  & \U           \\
\hline  $\mathrm{T}$        & $\mathrm{T}$  & \U            & $\mathrm{T}$ \\
\hline  $\mathrm{F}$        & \U            & $\mathrm{F}$  & $\mathrm{F}$ \\
\hline  \U                  & \U            & \U            & \U           \\
\hline
    \end{tabular}
}

\end{table}

The game has three possible outcomes: 
\true ($T$), 
\false ($F$),
and \unknown (\U), each of which carries its own reward structure.
Note that the game outcome is only used to determine rewards and penalties,
and does not necessarily correspond exactly to the oracle's output.
Indeed, anyone observing the system is free to compute oracle outcomes
as they wish and depending on the 
context of the proposition. For the sake of this presentation a
suggested mapping is presented in Table~\ref{tbl:outcomes}-b.
The suggested oracle output follows the voting outcome if it matches the
certification outcome or the certification outcome is \U.
The oracle is not restricted to an output of $\{T, F, \U\}$, but 
could instead have an output in the range $[0, 1]$ indicating confidence
in the truth or falsity of the proposition.

\begin{table*}[t]

\caption{Summary of rewards and penalties}
\label{tbl:rewards}

    \centering
    \begin{tabularx}{0.95\textwidth}{cYYYY}
        \cmidrule(lr){2-5}
          \multicolumn{1}{c}{\Yspace} 
        & \multicolumn{2}{c}{\textbf{Reward}} 
        & \multicolumn{2}{c}{\textbf{Penalty}} \\
        \cmidrule(lr){2-3} \cmidrule(lr){4-5} 
          \textbf{Outcome} 
        & \multicolumn{1}{Y}{\textbf{Voters}} 
        & \multicolumn{1}{Y}{\textbf{Certifiers}}
        & \multicolumn{1}{Y}{\textbf{Voters}}
        & \multicolumn{1}{Y}{\textbf{Certifiers}}\\
        \cmidrule(lr){1-1} 
        \cmidrule(lr){2-2} 
        \cmidrule(lr){3-3} 
        \cmidrule(lr){4-4} 
        \cmidrule(lr){5-5} 
            $\mathrm{T}$       
        &   $\mathrm{\left( \frac{s_{i,j,T}}{s_{TOT,j,T}} \right) \times B_j}$
        &   $\mathrm{\left( \frac{\sigma_{i,j,T}}{\sigma_{TOT,j,T}} \right) \times 
            \left( R_T \times \frac{1}{\tau} \right)}$ 
        &   $\mathrm{s_{i,j,F}}$
        &   $\mathrm{\sigma_{i,j,F}}$
        \\
     
            $\mathrm{F}$       
        &   $\mathrm{\left( \frac{s_{i,j,F}}{s_{TOT,j,F}} \right) \times B_j}$
        &   $\mathrm{\left( \frac{\sigma_{i,j,F}}{\sigma_{TOT,j,F}} \right) \times 
            \left( R_F \times \frac{1}{\tau} \right)}$ 
        &   $\mathrm{s_{i,j,T}}$
        &   $\mathrm{\sigma_{i,j,T}}$
        \\

             \O        
        &    $\mathrm{0}$    
        &    $\mathrm{0}$     
        &    $\mathrm{0}$     
        &    $\mathrm{\sigma_{i,j,F} + \sigma_{i,j,T}}$
        \\
        \cmidrule(lr){1-5}
    \end{tabularx}
\vspace*{-6pt}
\end{table*}

\subsection{Rewards and Penalties}

Broadly speaking, players are only rewarded for $T$ and $F$ outcomes in
which they took a position that matches it.
Conversely, those who took opposing positions are penalized.
In \unknown outcomes, certifiers are penalized and voters receive
no rewards or penalties. As argued in the paper, this scheme incentivizes the participants to
behave honestly on the validity of propositions.

For the rest of this subsection, fix a player $i \in [1,N]$ and
proposition $p_j \in P$ to be decided so as to
enumerate the rewards and penalties for each of the three possible
game outcomes.
Voter and certifier rewards are presented separately although, as noted earlier,
nothing prohibits a player from being both a voter and a certifier.

Rewards and penalties are distilled into a single value
$r_v$ for voting and $r_c$ for certification.
A negative value indicates a penalty, while a positive one indicates
a reward.
The results are summarized in Table~\ref{tbl:rewards}. 

\subsubsection{True and False Outcomes}

In the case of a $T$ outcome, the voting reward is as follows.

\begin{equation}\label{eq:tvote}
    r_v = 
    \left( \frac{s_{i,j,T}}{s_{TOT,j,T}} \right) \times B_j - 
    s_{i,j,F} 
\end{equation}

\noindent 
The player's vote reward is their share of the $T$-voting stake
times the proposition's bounty amount.
Their penalty is equal to their $F$-voting stake.
The certifier reward is shown in Eq.~\ref{eq:tcert} below.

\begin{equation}\label{eq:tcert}
    r_c = 
    \left( \frac{\sigma_{i,j,T}}{\sigma_{TOT,j,T}} \right) \times 
    \left( R_T \times \frac{1}{\tau} \right) -
    \sigma_{i,j,F} 
\end{equation}

A certifier's reward is equal to her share of the $T$-certifying
stake times the \textit{true certifier reward pool} amount $R_T$ times the
reciprocal of the \textit{certification target} $\tau$.
The true reward pool is a reward pool used to reward certifiers who
correctly certify articles as \true.
The certification target can be seen as the number of certifications that the 
pool should have enough funds to pay for. 
For instance, if $R_T = 1000$ and $\tau = 10$, then
$100$ monetary units will be distributed to the certifiers. 
The next proposition will have $R_T = 900$, and therefore $90$ units
will be distributed.

In the case of an $F$ outcome, the rewards and penalties are similar.

\subsubsection{Unknown Outcome}\label{section:oracle:unknown_outcome}

For an outcome of \U (\unknown), 
voters are neither rewarded nor penalized ($r_v = 0$),
while certifiers are penalized as follows.

\begin{equation}
    r_c = - \left( \sigma_{i,j,F} + \sigma_{i,j,T} \right) 
\end{equation}

That is, certifiers forfeit all of their stake, regardless of agreement
with voters.
The rationale for penalizing certifiers and not voters is that certifiers
chose to certify proposition $p_j$ while voters did not choose to vote on it
as they received it at random.

%
%

\subsection{Monetary Flows}

Note that voters are not rewarded for \unknown outcomes, and
therefore bounties are not claimed.
Thus far, the distribution of penalties and unclaimed bounties has not been
discussed, and the funding of reward pools and bounty amounts has only
been briefly mentioned.

Submitter fees are used to fund bounties, while the certifier reward pools are 
initially left empty.
Certifier reward pools are funded by unclaimed bounties and penalties.
In the absence of reward pools, certification will be rare, which will
lead to most bounties being unclaimed.
It is expected that this method will approach equilibrium 
where reward pools and bounties are balanced to provide reasonable
incentives for both voters and certifiers.
Further, it is adaptable as it can approach a new
equilibrium automatically if external conditions change.

An additional consideration is the draining of reward pools.
Each time a proposition is decided with voting outcome $T$,
an amount $\frac{R_T}{\tau}$ is subtracted from $R_T$.
If the certification outcome is also $T$, the funds are used to pay
out certifier rewards.
Otherwise, the funds are added to $R_F$.
A symmetric case applies for \false outcomes.
The reasoning behind this is explained in Section~\ref{section:bias} but
its intuition lies in the fact that 
 it ensures that the reward pools encourage certifiers to
certify equal numbers of \true and \false propositions, thereby discouraging
voters from voting with constant $T$ or $F$ values in order to 
maximize profit without considering the actual propositions.

\section{System Analysis}\label{section:analysis}

To analyze \Oracle we first construct formulas relating the parameters of the system
to the probability of the voting procedure producing incorrect results
({\em i.e.,} where the outcome for proposition $p_j$ with truth value $t_j$ is $\neg t_j$). 
Next, we introduce mathematics that relate the system parameters to the minimum
accuracy needed by voters so that they remain profitable.
Subsequently, we prove that a Nash equilibrium exists where all players
play honestly under a form of honest majority assumption on the voters.
Finally, we argue that the certifier reward structure avoids a situation
where players profitably vote and certify everything with a constant 
$T$ or $F$, even if the proposition list is heavily biased towards \true 
or \false propositions.

\subsection{Voting Outcomes and Manipulation}

This subsection determines the probability that voting 
outcomes are correct as a function of accuracy
and extends the analysis to determine an adversary's probability
of forcing incorrect outcomes.
For simplicity, we assume all non-adversarial players are honest, have the 
same accuracy $q$, and always vote with $s_{max}$ monetary units of stake.
We can therefore treat the voting process on a single
proposition as a sequence of $\frac{D_v}{s_{max}}$ Bernoulli trials with 
probability $q$ of success.
The probability that the voting outcome is correct is therefore:

\begin{equation*}
P\left[ B \left( \frac{D_v}{s_{max}}, q \right) > \frac{D_v}{2\cdot s_{max}} \right]
\end{equation*}

\noindent where $B(n,p)$ denotes a binomial random variable.
For example, if $D_v = 20$, $s_{max} = 1$, and $q = 0.8$,
the probability of obtaining a correct voting outcome is 
roughly $99.7$\%.

Now let us assume an adversary has $n$ monetary units and seeks
to force an incorrect outcome on a specific proposition.
For simplicity, we assume $n$ is a multiple of $s_{max}$ and that the 
proposition list does not change during the attack.
Each proposition can once again be modeled as a sequence of 
$\frac{D_v}{s_{max}}$ Bernoulli trials. 
Each trial is successful with probability: 

\begin{equation*}
p + (1-p)(1-q) = 1 - q + p \cdot q
\end{equation*}

\noindent where $p$ is the probability that the vote belongs to the adversary.

If the adversary uses all $n$ tokens to vote, then $\frac{n}{s_{max}}$
votes belong to the adversary across all $|P|$ propositions.
Once all propositions in the proposition list are decided, the probability
that an arbitrarily-chosen vote belongs to the adversary is:

\begin{equation*}
\frac{n}{s_{max}} \times \frac{s_{max}}{|P|\cdot D_v} = 
\frac{n}{|P| \cdot D_v}
\end{equation*}

\noindent So the probability that an arbitrary vote is incorrect is:

\begin{equation}\label{eq:adversary_control}
1 - q + \frac{n \cdot q}{|P| \cdot D_v}
\end{equation}

Note, both $|P|$ and $D_v$ appear in the denominator demonstrating that 
increasing these parameters renders system manipulation more difficult.
The probability of an adversary changing the outcome of a 
 proposition is shown below:

\begin{equation*}
P\left[ B \left( \frac{D_v}{s_{max}}, 1 - q + \frac{n\cdot q}{|P| \cdot D_v} \right) > \frac{D_v}{2\cdot s_{max}} \right]
\end{equation*}

It can be seen that if the quantity in Eq.~\ref{eq:adversary_control} 
is less than 0.5, there are parameter values that make it arbitrarily difficult 
for the adversary to force an incorrect result. 
Table~\ref{tbl:vote_manipulation} shows the adversary's chance of
success for various parameter values.
The six columns show the voting stake before decision,
the size of the proposition list,
the voter accuracy,
the fraction of votes controlled by the adversary,
the probability of forcing an incorrect output for a specific proposition,
and the probability of doing so for any proposition, respectively. 
The rows where no votes are controlled by an adversary
show the probability of an incorrect outcome occurring due to honest voter behavior alone.
It can be seen that if voters are 95\% accurate and $D_v = 100\cdot s_{max}$,
manipulation is effectively impossible, even by a powerful adversary controlling 
25\% of the votes.

\begin{table}
\caption{Probability of Adversary Manipulating Voting}
\label{tbl:vote_manipulation}
    \centering
	\setlength\tabcolsep{4pt}
    \begin{tabular}{|c|c|c|c|r|r|}
	\hline
      $\mathbf{D_v}$ 
    & $\mathbf{|P|}$
    & $\mathbf{q}$ 
    & $\mathbf{\frac{n}{|P|\cdot D_v}}$
    & $\mathbf{P(\textbf{Specific})}$ 
    & $\mathbf{P(\textbf{Any})}$
    \\ \hline
      $\mathrm{20 \cdot s_{max}}$
    & $\mathrm{100}$
    & $\mathrm{0.8}$
    & $\mathrm{0}$
    & $\mathrm{0.0006}$
    & $\mathrm{0.0548}$
	\\ \hline
      $\mathrm{20 \cdot s_{max}}$
    & $\mathrm{100}$
    & $\mathrm{0.8}$
    & $\mathrm{0.05}$
    & $\mathrm{0.0028}$
    & $\mathrm{0.2438}$
	\\ \hline
      $\mathrm{20 \cdot s_{max}}$
    & $\mathrm{100}$
    & $\mathrm{0.8}$
    & $\mathrm{0.25}$
    & $\mathrm{0.1275}$
    & $\mathrm{\approx 1}$
	\\ \hline
      $\mathrm{20 \cdot s_{max}}$
    & $\mathrm{100}$
    & $\mathrm{0.95}$
    & $\mathrm{0}$
    & $\mathrm{\lowp{-9}}$
    & $\mathrm{\lowp{-7}}$
	\\ \hline
      $\mathrm{20 \cdot s_{max}}$
    & $\mathrm{100}$
    & $\mathrm{0.95}$
    & $\mathrm{0.05}$
    & $\mathrm{\lowp{-6}}$
    & $\mathrm{\lowp{-4}}$
	\\ \hline
      $\mathrm{20 \cdot s_{max}}$
    & $\mathrm{100}$
    & $\mathrm{0.95}$
    & $\mathrm{0.25}$
    & $\mathrm{0.0123}$
    & $\mathrm{0.7100}$
	\\ \hline
      $\mathrm{100 \cdot s_{max}}$
    & $\mathrm{100}$
    & $\mathrm{0.8}$
    & $\mathrm{0}$
    & $\mathrm{\lowp{-11}}$
    & $\mathrm{\lowp{-9}}$
	\\ \hline
      $\mathrm{100 \cdot s_{max}}$
    & $\mathrm{100}$
    & $\mathrm{0.8}$
    & $\mathrm{0.05}$
    & $\mathrm{\lowp{-8}}$
    & $\mathrm{\lowp{-6}}$
	\\ \hline
      $\mathrm{100 \cdot s_{max}}$
    & $\mathrm{100}$
    & $\mathrm{0.8}$
    & $\mathrm{0.25}$
    & $\mathrm{0.0168}$
    & $\mathrm{0.8156}$
	\\ \hline
      $\mathrm{100 \cdot s_{max}}$
    & $\mathrm{100}$
    & $\mathrm{0.95}$
    & $\mathrm{0}$
    & $\mathrm{\approx 0}$
    & $\mathrm{\approx 0}$
	\\ \hline
      $\mathrm{100 \cdot s_{max}}$
    & $\mathrm{100}$
    & $\mathrm{0.95}$
    & $\mathrm{0.05}$
    & $\mathrm{\approx 0}$
    & $\mathrm{\approx 0}$
	\\ \hline
      $\mathrm{100 \cdot s_{max}}$
    & $\mathrm{100}$
    & $\mathrm{0.95}$
    & $\mathrm{0.25}$
    & $\mathrm{\lowp{-5}}$
    & $\mathrm{0.0002}$
	\\ \hline
    \end{tabular}
    
    \vspace*{-6pt}
\end{table}

\subsection{Minimum Voting Accuracy}\label{section:minimum_accuracy} 

According to the analysis in the previous subsection, the accuracy of honest
voters is critical to the security of \Oracle.
This subsection quantifies the minimum accuracy 
(\textit{i.e.}, the probability that the player's belief $\beta_i(p_j)$ 
matches the truth value $t_j$ of proposition $p_j$)
players need to achieve profitability in expectation
by constructing formulas that relate the accuracy threshold 
to the parameters of the system. 
Since making it more difficult to earn a profit is expected to lower
participation, it is critical to set the parameters carefully to achieve
a reasonable tradeoff between accuracy and participation.

For simplicity, we assume the parameters are set such that the probability
of incorrect decisions is negligible.
Consider player $i$ with accuracy $q_i$. 
A vote with stake of $s_{max}$ on proposition $p_j$ yields a profit in 
expectation when, at decision time for $p_j$:

\begin{equation}\label{eq:vprofit}
\left[ \frac{q_i \cdot s_{max}}{ \max(s_{TOT,j,T}, s_{TOT,j,F}) } 
\right] B_j > 
(1-q_i) \cdot s_{max}
\end{equation}

In other words, the expected share of the voting rewards is greater 
than the expected penalties.
Note that there is no need to account for \U outcomes since voters
receive neither rewards nor penalties in that case.
At decision time, the denominator is clearly at least half of $D_v$, and at 
most $D_v$. 
Therefore: 

\begin{equation}\label{eq:shares}
\frac{1}{2} \cdot D_v \le 
\max( s_{TOT,j,T}, s_{TOT,j,F} ) \le 
D_v 
\end{equation}

It is clear from Eq.~\ref{eq:vprofit} that the voter is profitable
for sufficiently high values of $B_j$.
Towards the goal of computing an upper bound, 
we over-approximate the range in which a voter is profitable 
using Eq.~\ref{eq:shares} as follows.

\begin{equation*}
\left( \frac{q_i \cdot s_{max}}{ D_v } 
\right) B_j > 
(1 - q_i) \cdot s_{max}
\end{equation*}

\noindent Re-arranging yields the following upper bound on $B_j$:

\begin{equation}\label{eq:bcap}
B_j \le \frac{(1 - q_i) \cdot D_v}{q_i}
\end{equation}

By capping the bounty according to Eq.~\ref{eq:bcap}, 
it is possible to enforce a minimum accuracy such that below
that threshold voting becomes 
unprofitable.
For instance, if 80\% accuracy is desired
and $D_v = 1000$, the bounty must be capped at $250$ monetary units.
If instead 50\% accuracy is desired the bounty must be capped at $1000$. 
Of course this analysis only lower-bounds the threshold; it
neglects the voter's costs in terms of time to evaluate propositions,
computing power, and blockchain transaction fees.

\subsection{Desirable Nash Equilibrium}

This subsection demonstrates the existence of a desirable Nash equilibrium
in which all players are honest.
Indeed, it is clear that any situation in which all players always
vote and certify in concert is a Nash equilibrium, 
as any player who votes against all the others will only stand to lose.
However, we seek to show that such an equilibrium exists under the
assumption that the quantity in Eq.~\ref{eq:adversary_control} is less than 
$0.5$.
In effect, this assumes that honest voters are sufficiently accurate and
in such plurality that a majority of votes are correct.
From the analysis in Section~\ref{section:minimum_accuracy}, the assumption is
sufficient to show that there exists an assignment to the parameters such that 
all propositions have the correct voting outcome with overwhelming probability.

While we omit the details of the analysis due to space requirements, 
it is clear that playing honestly is a Nash equilibrium when every voting outcome 
is correct and only feasible strategies are considered.
Rewards are only ever paid to players who agree with the voting outcome.
Since every proposition $p_j$ has the voting outcome $t_j$, if player $i$ votes
or certifies $\beta_i(p_j)$ (\textit{i.e.}, honestly),
she agrees with the voting outcome with probability $q_i$.
Naturally, an honest player could perform better by switching to a ``perfect''
strategy in which they always vote correctly rather than honestly,
but such a strategy is not feasible since players do not know the underlying truth values.
Since all beliefs are independent by assumption, a player $p_i$ cannot develop
complex strategies that leverage statistical correlation between beliefs
to vote correctly with probability better than $q_i$.
Even an adversary with perfect accuracy who controls 100\% of the certifying 
stake is incentivized to play honestly (or not play at all \textit{e.g.}, 
in the case where $R_T = 0$ and every proposition is \true).
Indeed, any other strategy results in the adversary losing \textit{all} 
of their stake.

\subsection{Proposition Bias and Reward Pools}\label{section:bias}

While the previous subsection demonstrates that playing honestly is an
equilibrium under the assumption that an adversary does not control voting,
this assumption may not hold if a dishonest strategy is easier and still
profitable.
In this subsection, we identify a candidate for such a strategy and argue that 
the reward structure combats it.

Imagine \true propositions are more common than \false ones
and define as a \textit{lazy voter} one who always votes $T$.
Let $p = P(t_j = T)$ denote the probability that a random proposition is \true.
We further assume as before that all voting outcomes are correct, so
the lazy voter agrees with the voting outcome on every \true proposition 
and disagrees on every \false proposition.
Intuitively, this lazy strategy seems viable, but note that
it is also necessary to consider certifier behavior since
without certification no rewards are paid out.
Certifier incentives are tied to the certifier reward pools, and their values
fluctuate over time.

When a \true proposition is decided, the $R_T$ pool will shrink by
$\frac{R_T}{\tau}$.
Over time the $R_T$ pool will drain much faster than the $R_F$
pool when $p > 0.5$.
Indeed, the $R_F$ pool may actually grow in this case.

Again, we omit the details of the formal argument due to space restrictions. 
However, informally, this process incentivizes certifiers
to make more certifications on propositions that they believe are \false,
since the potential rewards are greater.
At equilibrium, roughly equal amounts of \true and \false propositions
should carry certifications, meaning that the lazy strategy will not be
profitable.
Note that certifiers are never incentivized to certify a \true proposition
as \false in an attempt to acquire the $R_F$ reward, as in that case the
certifier will disagree with voters and be penalized.

\section{Extension to Unknown Propositions and Data Availability}\label{section:extension}

This section presents an extension to simultaneously handle two issues that would arise from a practical implementation of \Oracle.

\begin{enumerate}
\item
A proposition $p$ may not have a clear answer between \true and \false, \textit{e.g.} if the proposition is ``Is complexity class
P equal to NP?''~\cite{p_vs_np}
Since voters are assigned a proposition randomly, they may be assigned a proposition for which they cannot vote towards the truth value, since the truth for some is unknown.

\item
As storage directly on the blockchain is relatively expensive~\cite{blockchain_rent}, an implementation of \Oracle would most likely use an off-chain storage solution for propositions, allowing them to be uniquely identified by an immutable public hash~\cite{multihash}. As such, the
problem of data availability~\cite{data_availability} presents itself: what if the submitter does not propagate her proposition sufficiently for the majority of the network to see it, or a malicious actor submits a proposition but intentionally does not share the data corresponding to the hash?
Voters would again be unable to vote towards the truth, in this case because the question is unavailable to them.
\end{enumerate}

In both cases, the assumption in Section~\ref{section:preliminaries:assumptions} that each proposition has a truth value is violated.
\Oracle can be extended by adding a third option for voters, this of \unknown.
If the voting outcome is \unknown (\U) due to a majority of voters voting \unknown, voters are neither rewarded nor punished, as in Section~\ref{section:oracle:unknown_outcome}. 
Certifiers may not take a position of \unknown, they should simply abstain from participating in such cases.
The reward structure for the game outcomes remains the same as described in Section~\ref{section:oracle}, and the analysis of desiderata follows in a straightforward manner from that in Section~\ref{section:analysis}.
This extension protects players from penalization when submitters provide unclear
propositions or withhold data.

\section{Applications}\label{section:applications}

Practical applications of a decentralized, permissionless, and trustless oracle are numerous.
This section gives an overview of candidate use cases for \Oracle.

\smallskip \noindent {\bf Machine Learning and Data Annotation:}
Modern data science techniques require a tremendous amount of annotated data~\cite{data_annotation} to train predictive models, with applications ranging from self-driving cars, to targeted marketing or medical diagnoses.
Centralized crowd-sourcing platforms~\cite{crowdflower,mturk} offer a means for individuals to perform human intelligence tasks and receive a payment.
However, there have been reports of these platforms not compensating workers properly~\cite{IEEE:6686081} and due to the lack of incentives to label data correctly, low-quality labeling is prevalent in the industry.
This requires customers to pay for redundant work, since there is no easy means of actually determining which labels are incorrect.
A decentralized oracle that incentivizes workers to vote honestly can be used for data annotation, potentially reducing cost and improving quality over existing solutions.

\smallskip
\noindent {\bf Data Availability:}
Decentralized applications that make use of off-chain resources~\cite{truebit} suffer from the data availability problem~\cite{data_availability}, in which off-chain data may only exist transiently, or may never have existed at all (\textit{e.g.} in the case of a malicious actor).
\Oracle, with the extension of Section~\ref{section:extension}, can be used as a data availability oracle for all such applications.

\smallskip
\noindent {\bf Adjudication Mechanisms:}
Negotiations between parties that require an adjudication mechanism can instead use a decentralized oracle such as \Oracle.
In this case, the oracle will essentially serve as a public jury.
Decentralized applications that deal with real-world resources, such as legal agreements, transfers of assets, etc. can make use of this system.

\section{Conclusion}\label{section:conclusion}

This work introduces a decentralized, trustless, and permissionless blockchain oracle system, \Oracle.
Submitters enter propositions into the system, while voters and certifiers play a game to determine the truth value of each proposition.
We analyze the game-theoretical incentive structure to show that a desirable Nash equilibrium exists whereby under
a set of simple assumptions all rational players behave honestly.


\bibliographystyle{IEEEtran}
\bibliography{refs}
 
\end{document}